\documentclass[useAMS, subeqn, usenatbib]{mn2e}
\usepackage{epsf,times}
\usepackage{natbib}
\usepackage{graphicx}
\usepackage{epsfig}
\usepackage{hyperref}
\usepackage{amsmath}
\usepackage{amssymb}
\usepackage{wasysym}
\usepackage{textgreek}
\usepackage{color}
\usepackage{pifont}
\usepackage{comment}
\usepackage{tablefootnote}
\usepackage{datetime}

\newdateformat{monthyeardate}{%
  \monthname[\THEMONTH], \THEYEAR}

\title[\textbeta~Cir~B]{Discovery of a brown dwarf companion to the A3V star \textbf{\textbeta}~Circini}
\date{\monthyeardate\today}
\author[L.C. Smith et al.]{L.C. Smith$^1$\thanks{L.SMITH10@herts.ac.uk}, P.W. Lucas$^1$, C. Contreras~Pe\~{n}a$^{2,1,4}$, R. Kurtev$^{3,4}$, F. Marocco$^1$, H.R.A. Jones$^1$,
\newauthor
J.C. Beamin$^{5,10,4}$, R. Napiwotzki$^1$, J. Borissova$^{3,4}$, B. Burningham$^{1,6}$, J. Faherty$^{7,8}$\thanks{Hubble Fellow}, 
\newauthor
D.J. Pinfield$^1$, M. Gromadzki$^{4,3}$, V.D. Ivanov$^{9,10}$, D. Minniti$^{2,11}$, W. Stimson$^1$, V. Villanueva$^{4,3}$\\
$^1$ Centre for Astrophysics Research, Science and Technology Research Institute, University of Hertfordshire, Hatfield AL10 9AB, UK\\
$^2$ Departamento de Ciencias Fisicas, Universidad Andres Bello, Republica 220, Santiago, Chile\\
$^3$ Istitiuto de F\'{i}sica y Astronom\'{i}a, Universidad de Valpara\'{i}so, ave. Gran Breta\~{n}a, 1111, Casilla 5030, Valpara\'{i}so, Chile\\
$^4$ Millennium Institute of Astrophysics, Av. Vicua Mackenna 4860, 782-0436, Macul, Santiago, Chile\\
$^5$ Instituto de Astrof\'{i}sica, Facultad de F\'{i}sica, Pontificia Universidad Cat\'{o}lica de Chile, Casilla 306, Santiago 22, Chile\\
$^6$ NASA Ames Research Center, Mail Stop 245-3, Moffett Field, CA 94035, USA\\
$^7$ Department of Terrestrial Magnetism, Carnegie Institution of Washington, Washington, DC 20015, USA\\
$^8$ Department of Astrophysics, American Museum of Natural History, Central Park West at 79th Street, New York, NY 10034\\
$^9$ European Southern Observatory, Karl-Schwarzschild-Strasse 2, D-85748 Garching bei Munchen, Germany\\
$^{10}$ European Southern Observatory, Ave. Alonso de Cordova 3107, Vitacura, Santiago 19001, Chile\\
$^{11}$ Vatican Observatory, V00120 Vatican City State, Italy\\
}

\begin{document}

\maketitle

\begin{abstract}
We report the discovery of an L dwarf companion to the A3V star \textbeta~Circini. VVV~J151721.49-585131.5, or \textbeta~Cir~B, was identified in a proper motion and parallax catalogue of the Vista Variables in the V\'{i}a L\'{a}ctea survey as having near infrared luminosity and colour indicative of an early L dwarf, and a proper motion and parallax consistent with that of \textbeta~Cir. The projected separation of $\sim$3.6$\arcmin$ corresponds to $6656$~au, which is unusually wide. The most recent published estimate of the age of the primary combined with our own estimate based on newer isochrones yields an age of $370-500$~Myr. The system therefore serves as a useful benchmark at an age greater than that of the Pleiades brown dwarfs and most other young L dwarf benchmarks. We have obtained a medium resolution echelle spectrum of the companion which indicates a spectral type of L1.0$\pm$0.5 and lacks the typical signatures of low surface gravity seen in younger brown dwarfs. This suggests that signs of low surface gravity disappear from the spectra of early L dwarfs by an age of $\sim370-500$~Myr, as expected from theoretical isochrones. The mass of \textbeta~Cir~B is estimated from the BHAC15 isochrones as $0.056\pm0.007$~M$_{\odot}$.
\end{abstract}

\begin{keywords}
stars: individual: \textbeta~Circini - binaries:general - brown dwarfs
\end{keywords}

\section{Introduction}

Source confusion in the Galactic plane has meant that many of the nearest stars have gone unnoticed until relatively recently. To find even relatively nearby brown dwarfs one must look in the infrared, but here the problem of source confusion is greater still. Furthermore, in the Galactic plane there is a degeneracy between spectral type and interstellar extinction that makes it difficult to use broad band colour selections to distinguish nearby L dwarfs from reddened normal stars, unless proper motion data are available \citep{folkes12}. A search of recent literature (see \citealt{smith14b} and references therein) shows that the number of known brown dwarfs in the Galactic plane has increased dramatically in the last decade, but there is still much scope for further discovery.

The well known degeneracy between brown dwarf ages and masses means that neither of these properties are well constrained through observational methods alone, yet they are essential to understanding the mass function of the local field. The masses of early L type dwarfs (T$_{\text{eff}}\sim$2000~K) in particular span a range which includes hydrogen burning stars (L dwarfs older than a few Gyr) and planetary mass objects (L dwarfs younger than a few tens of Myr).

Members of gravitationally bound systems serve as useful benchmark objects when a component of the system has well constrained attributes (eg. age, metallicity). As members of such systems can be assumed to have formed from the same molecular cloud at a similar time, the same attributes can be adopted for all members of the system, see e.g. \citet{pinfield06}, \citet{deacon14}.

Identification of benchmark brown dwarfs across a range of effective temperatures, ages, and metallicities is necessary to refine atmospheric models to the point where they can accurately reproduce observed spectra.

In this paper, we report the identification of a widely separated L1 type brown dwarf companion to the A3V type main sequence star \textbeta~Circini (HD~135379, HIP~74824). This object adds to the small number of brown dwarf companions to early type stars previously known: only 8 such systems are listed by \citet{derosa14b}. 
The primary allows us to place a useful age constraint on the system, but does not constrain the metallicity since abundance measurements of A type stars are problematic \citep{adelman07}. 
In Section \ref{astrometry} we detail our astrometric measurements and consider the binarity of the pair. Section \ref{age} deals with the age of the system. In Section \ref{spectrum} we describe our spectroscopic observation of the secondary, measure a spectral type and radial velocity, and investigate the gravity-sensitive spectral features. In Section \ref{discussion} we discuss how this discovery compares to other brown dwarf age benchmarks.

\section{Astrometry and Companionship}\label{astrometry}
\begin{figure}
\begin{center}
\begin{tabular}{c}
\epsfig{file=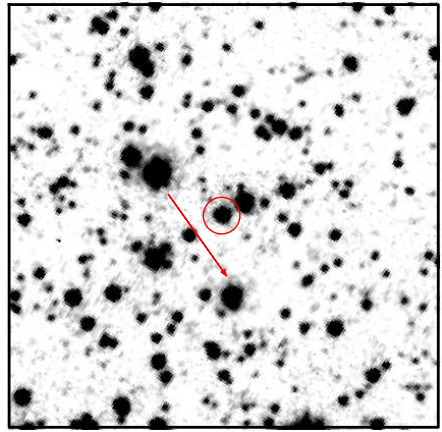,width=1\linewidth,clip=}
\end{tabular}
\caption{A $1$\arcmin$\times{}1$\arcmin~VVV $K_{\text{s}}$ band image of \textbeta~Cir~B. North is up and east is to the left. The arrow shows the direction of motion.}
\label{fig:skyplot}
\end{center}
\end{figure}

The Vista Variables in the V\'{i}a L\'{a}ctea (VVV, \citealt{minniti10}) is a public ESO near-infrared time domain survey of the southern Galactic plane and bulge using the wide-field camera VIRCAM on the 4m VISTA telescope at Paranal observatory in Chile. Pipeline data reduction, catalogue generation and calibration of the photometry and astrometry for VISTA data are provided by CASU and described by \citet{lewis10}.  and Smith et al. (in prep). As part of an early search for new brown dwarf benchmark objects we cross-matched preliminary VVV colour and luminosity selected brown dwarf candidates from \citet{smith15} to the Hipparcos astrometric catalogue. Matches were required to be within $300$\arcsec and also have reasonably consistent proper motion. We identified VVV~J151721.49-585131.5 (2012.0 designation, see Figure \ref{fig:skyplot}) as a match to \textbeta~Circini and we refer to it hereafter as  \textbeta~Cir~B. The object was also identified independently in a separate search for high proper motion objects in the first two years of VVV data by members of our team, (Kurtev et al., in prep).

Our astrometric pipeline is largely automated but here we describe our method below as it applies to \textbeta~Cir~B.

\begin{figure}
\begin{center}
\begin{tabular}{c}
\epsfig{file=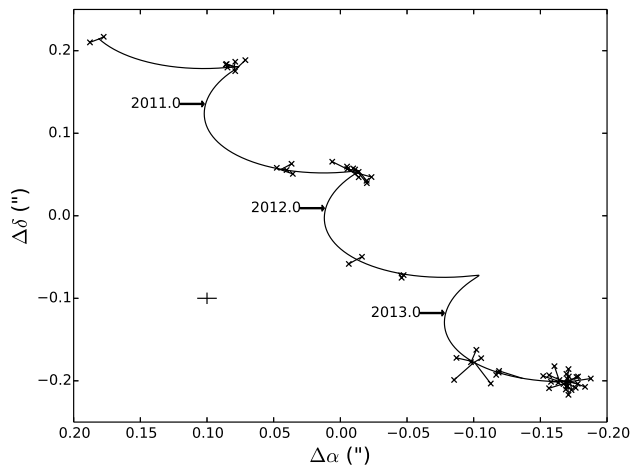,width=1\linewidth,clip=}
\end{tabular}
\caption{The fit of the parallax and proper motion (solid curve) to the position of \textbeta~Cir~B at each of the VVV epoch (black crosses). Solid lines join each observed position to the corresponding epoch on the parallax and proper motion curve. A representative single epoch error bar is shown. Epochs 2011.0, 2012.0, and 2013.0 are indicated for reference. The motion of \textbeta~Cir~B is characterised by a relative proper motion of $-90.1\pm1.0$ and $-126.5\pm1.2$~mas~yr$^{-1}$ in $\alpha \cos{\delta}$ and $\delta$ respectively and a parallax of $33.6\pm1.6$~mas.}
\label{fig:pmplx_fit}
\end{center}
\end{figure}

We used the 58 Ks bandpass pawprint catalogues (with seeing $<1.2$\arcsec of VVV tile d018 available to us as of April 30th 2014). We select references sources from within a radius sufficient to ensure at least 50 good quality stars are in each quadrant surrounding the target, in this case $93$\arcsec. A further iterative coordinate transformation and rejection of reference sources with significant proper motion resulted in a final selection of 142 reference sources. Fitting of the proper motion and parallax in both the $\alpha \cos{\delta}$ and $\delta$ dimensions was performed in \textsc{matlab} using a robust technique involving an iterative reweighting of data points as a function of their residuals. We measure proper motions of $-90.1\pm1.0$ and $-126.5\pm1.2$~mas~yr$^{-1}$ in $\alpha \cos{\delta}$ and $\delta$ respectively, and parallaxes of $33.1\pm1.7$ and $35.7\pm3.5$~mas in $\alpha \cos{\delta}$ and $\delta$. The inverse variance weighted average of these two measurements is the overall parallax, $33.6\pm1.6$~mas ($29.8\pm1.4$~pc).
Figure \ref{fig:pmplx_fit} shows our astrometric fit over the VVV positional data.

The Hipparcos proper motion for \textbeta~Cir~A (saturated in VVV) is $-97.74\pm0.28$ and $-134.15\pm0.22$~mas~yr$^{-1}$ in $\alpha \cos \delta$ and $\delta$ respectively. The parallax is $32.73\pm0.19$~mas ($30.55\pm0.18$~pc) \citep{vanleeuwen07}. \citet{kharchenko07} provide a radial velocity of $+9.6\pm1.8$~km~s$^{-1}$.

The apparent slight disagreement in proper motion ($\Delta\mu=10.8\pm1.6$~mas~yr$^{-1}$) of the two objects can be accounted for by a translation of the VVV proper motion to an absolute reference frame. The mean offset in the motion of nearby ($<3$\arcmin) PPMXL sources common to our catalogue is significant at $-6.6\pm2.3$~mas~yr$^{-1}$ in both $\alpha \cos \delta$ and $\delta$ (the medians are $-8.5$ and $-4.0$~mas~yr$^{-1}$ respectively). 
Factoring in this offset brings the overall proper motion of \textbeta~Cir~B to $-96.7\pm2.5$ and $-133.1\pm2.6$~mas~yr$^{-1}$ in $\alpha \cos \delta$ and $\delta$ and the proper motion difference between the pair to $\Delta\mu=1.5\pm3.6$~mas~yr$^{-1}$.
The angular separation between the pair is $217.8$\arcsec, which gives a projected separation of $6656\pm40$~au using the Hipparcos distance of the primary. Given that the parallaxes, proper motions, and radial velocities (see Section \ref{RV}) of the pair are in close agreement, this establishes \textbeta~Cir~B as a genuine wide binary companion. We note that in the case of a face-on, circular orbit and adopting a mass of $2$~M$_{\odot}$ for \textbeta~Cir~A, the orbital velocity of \textbeta~Cir~B would present as a $\sim$3.6~mas~yr$^{-1}$ proper motion relative to \textbeta~Cir~A.

In an earlier study, \textbeta~Cir~A was a common proper motion companion search target of the VAST survey (\citealt{derosa14a}, separation coverage $4-45$~kau). The companion was missed because it does not appear in the proper motion catalogues queried, which required an optical detection. Separately, the \citet{ivanov13} visual search of the VVV images for new companions to known high proper motion stars did not identify \textbeta~Cir~B as the primary did not meet their $\mu>200$~mas~yr$^{-1}$ selection criterion.

\subsection{Lower Mass Companions}

The VVV proper motion catalogues for this tile have a high completeness out to $K_{\text{s}}=16.5$~mag, and no further companions were identified within them. Their completeness reduces with increasing proper motion, but the proper motion of \textbeta~Cir~A is relatively small in this regard.

We searched for fainter, even less massive companions by separately stacking 5 high quality VVV tile images from 2010 (which included the deeper master image taken for each field at the start of the survey) and 10 images from 2014-2015 and blinking them. No object within a 1'$\times$1' ($\sim1800$~au) field around \textbeta~Cir~B with a similar proper motion was detected down $K_{\text{s}}=18.0$~mag, which is the magnitude of the faintest nearby objects for which we are confident we would be able to identify a positional shift during blinking. Brown dwarf pairs are not found with separations over a few tens of au (e.g. \citealt{dupuy11}), presumably because of their low binding energy. Therefore, it is quite likely that any faint companion to \textbeta~Cir~B would be unresolved in the VVV images.
Using the Hipparcos parallax of the primary and the magnitude vs. spectral type relation of \citet{dupuy12} we determine that the $K_{\text{s}}=18.0$~mag upper limit corresponds to a spectral type of approximately T7.



\section{The Age of the \textbf{\textbeta}~Cir~System}\label{age}
	We can place an initial upper limit on the age of \textbeta~Cir~A equal to the main sequence lifetime. For a A3V type $\sim$2~M$_{\odot}$ star this is $\sim$1.25~Gyr based on the solar metallicity models of \citet{bressan12}. 

	There are values of the age of \textbeta~Cir~A in the literature: \citet{lachaume99} derive an age of $245\substack{+110 \\ -119}$~Myr based on \citet{bertelli94} isochrones; \citet{song01} derive an age of $166\substack{+212 \\ -116}$~Myr based on Str\"{o}mgren $uvby\beta$ photometrically determined effective temperature, surface gravity and metallicity and \citet{schaller92} theoretical evolutionary tracks; and \citet{nielsen13} used the \citet{siess00} tracks to give a $68\%$ confidence interval between $272$ and $458$~Myr with a median at $367$~Myr using Bayesian inference to determine the relative likelihoods of combinations of mass, age and metallicity. It has been shown that ages produced using more recent isochrones are generally older than previously (see e.g. \citealt{mamajek12}). Hence the greater age given by \citet{nielsen13} than previous estimates (though at the upper limit of their uncertainties) is unsurprising.
	
	The most accurate method of age dating an early type star is through association with a companion or system with a known age (e.g. a young moving group, stellar cluster). \citet{nakajima12} conclude that \textbeta~Cir~may be a member of the TW Hydrae stellar kinematic group ($\sim8-12$~Myr) and three other moving groups at a lower probability. The BANYAN II web tool (\citealt{gagne14}, \citealt{malo13}) operating under the assumption that \textbeta~Cir~is $<1$~Gyr old gives a zero percent probability of TW Hydrae membership and 66, 21 and 13 percent probabilities of membership to the \textbeta~Pictoris ($\sim12-22$~Myr), AB Doradus ($\sim70-120$~Myr) and field groups respectively.
	Indicators of youth in addition to spatial and kinematic agreement with a moving group are required to consider an object a genuine member. Membership to the TW Hydrae, \textbeta~Pictoris and AB Doradus moving groups are effectively ruled out due to incompatibility with the isochronal age.
    
\begin{figure}
\begin{center}
\begin{tabular}{c}
\epsfig{file=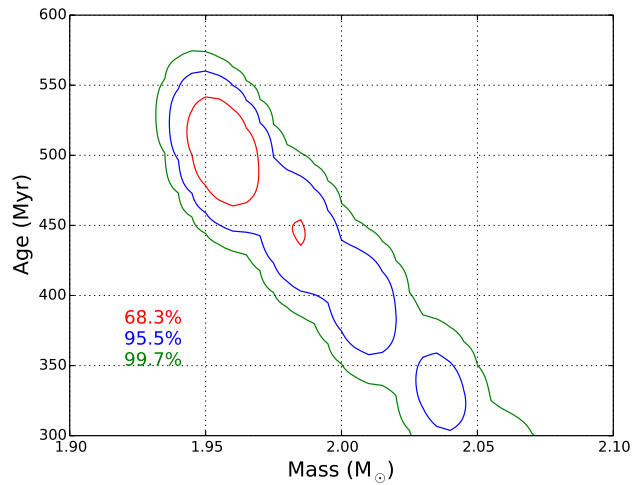,width=1\linewidth,clip=}
\end{tabular}
\caption{The age and mass probability distribution of \textbeta~Cir~A using \citet{bressan12} isochrones.}
\label{fig:bayes_plot}
\end{center}
\end{figure}

\begin{figure*}
\begin{center}
\begin{tabular}{c}
\epsfig{file=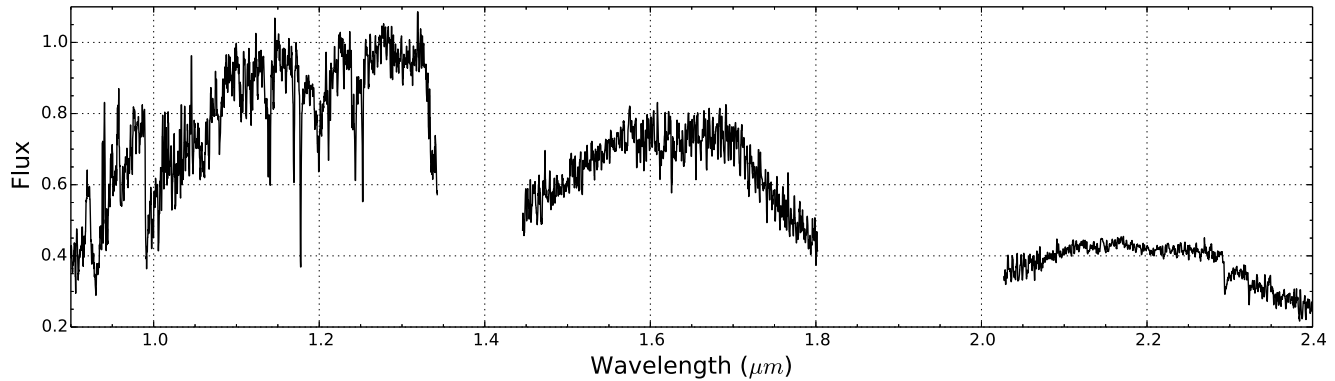,width=1\linewidth,clip=}
\end{tabular}
\caption{Our FIRE spectrum of \textbeta~Cir~B. The $Y$ and $J$ bands of the spectrum have been normalised to unity at $1.28$~\textmu{}m. Due to difficulty in accurately merging the echelle orders across the low signal to noise region between the $J$ and $H$ bands, the $H$ and $K$ bands of the spectrum have been normalised to the L1 standard using the flux measured between $1.65$~\textmu{}m and $1.75$~\textmu{}m.}
\label{fig:spectrum}
\end{center}
\end{figure*}
	
    For our own estimate of the age and mass of \textbeta~Cir~A we adopted a Bayesian maximum likelihood approach similar to that described by \citet{nielsen13} but using the more recent \citet{bressan12} isochrones. We used the mean effective temperature of \textbeta~Cir~A from seven literature sources (T$_{\text{eff}}=8676\pm33$~K), after discarding one significant outlier. The uncertainty on this measurement we have taken as the standard error on the mean. The Hipparcos parallax and Tycho-2 $V$ magnitude \citep{hog00} give us an absolute Tycho $V$ magnitude M$_{V}=1.65\pm0.02$. 
    We adopted a flat prior probability on the age, a Salpeter mass function and a normal probability distribution on [M/H] with a mean of $0.0$~dex and standard deviation of $0.1$~dex. This metallicity distirbution is suitable for young stellar populations in the solar neighborhood see e.g. \citep{nieva12}. 
    We treat metallicity as a floating parameter in spite of existing measurements ([Fe/H]~$\sim0.20$~dex, \citealt{erspamer03}) since measurements in the photosphere of an A type star cannot be assumed to reflect the true abundances of the star (and also in our case the companion). This is due to processes such as radiative diffusion \citep{adelman07}. This is apparent in the case of \textbeta~Cir~A from the wide scatter in individual elemental abundances \citep{erspamer03}. 
    We sampled metallicity in $0.05$~dex increments between $-0.3$ and $0.3$~dex, mass in $0.001$~M$_{\odot}$ increments between $1.7$ and $2.3$~M$_{\odot}$, and age in $1$~Myr increments between $200$ and $800$~Myr.
    This approach yielded a median age of $495$~Myr with a 68\% confidence interval of $442$ to $519$~Myr, 
    a median mass of $1.958$~M$_{\odot}$ with a 68\% confidence interval of $1.952$ to $1.986$~M$_{\odot}$, 
    and a median [M/H] of $0.0$~dex with a 68\% confidence interval of $-0.1$ to $0.1$~dex (i.e. essentially the same as the metallicity prior probability distribution). 
    Figure \ref{fig:bayes_plot} shows the age and mass probability distribution.

	The discrepancy between ages based on \citet{siess00} and \citet{bressan12} isochrones has been noted before \citep{derosa14b}. This is probably due to subtle differences in their input physics (e.g. the mixing length parameter, see \citealt{bell13}). We also note that \citet{siess00} used an earlier value for the solar metallicity (Z~$=0.02$) than \citealt{bressan12} (who used Z~$=0.0152$). However, we find that this latter difference has little effect on our derived age if we change our prior distribution for metallicity, given that the metallicity is allowed to vary.
In keeping with \citet{derosa14b} we adopt an age which incorporates the two isochronal ages, $370$ to $500$~Myr.

    Several studies have found evidence of a warm disk around \textbeta~Cir~A (\citealt{morales09}, \citealt{mcdonald12}, \citealt{ballering13}, \citealt{chen14}). Such disks can endure for $\sim1$~Gyr (\citealt{chen14}), so this does not constrain the system's age any better than the main sequence lifetime for an A3V star. \textbeta~Cir~B is also unlikely to interact with such a disk unless on a highly elliptical orbit given the current physical separation of at least $6656$~au.

\section{The Spectrum and Properties of \textbf{\textbeta}~Cir~B}\label{spectrum}

\subsection{Observation}
We observed \textbeta~Cir~B with the Folded-port InfraRed Echellette (FIRE) spectrograph in echelle mode (R$\sim6000$) on the Magellan Baade telescope at Las Campanas Observatory on the 26th of April 2015. The wavelength coverage of the instrument as configured is $0.85$ to $2.5$~\textmu{}m. The observation consisted of 4 integrations in an ABBA nodding pattern and we observed the A0 star HIP 76244 as a telluric standard. The slit was oriented 45$^{\circ}$ east of north to avoid contamination by background stars. The integration time for \textbeta~Cir~B was $253.6$~s in each position and observing conditions were good. The FIREHOSE pipeline was used to reduce and calibrate each order in the target and the standard. The profile of the object and background sky residuals are fitted using an iterative procedure whereby a bspline function is used for orders with high signal-to-noise and a Gaussian profile is assumed for low signal-to-noise orders. The pipeline then performs optimally weighted extraction of the object using the profile determined previously. The final spectrum is combined by scaling overlapping orders. Figure \ref{fig:spectrum} shows the reduced spectrum, the average signal-to-noise ratios in the Y, J, H and K bands are approximately 9, 22, 36 and 26 respectively per spectral pixel at R$\sim6000$.

\subsection{Spectral Type}\label{spty}
\begin{figure}
\begin{center}
\begin{tabular}{c}
\epsfig{file=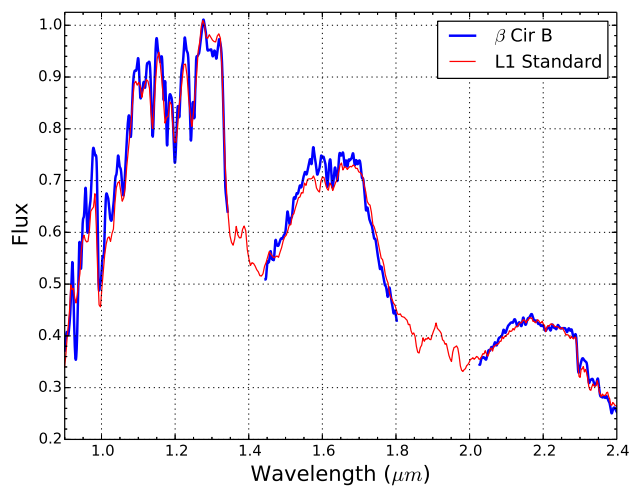,width=1\linewidth,clip=}
\end{tabular}
\caption{In blue, the spectrum of \textbeta~Cir~B smoothed to approximate the resolution of the SpeX data of the spectral standard. In red are the SpeX data of the L1.0 spectral standard 2MASSW J2130446-084520. Flux normalisation was performed using the same procedure as for Figure \ref{fig:spectrum}}
\label{fig:L1_comparison}
\end{center}
\end{figure}

We compared the reduced spectrum of \textbeta~Cir~B to the M and L dwarf standards of the SpeX Prism Spectral Libraries both visually and through a best-fit analysis and find the most similar template to be the near infrared L1.0 standard 2MASSW J2130446-084520 (\citealt{kirkpatrick10}, see Figure \ref{fig:L1_comparison}). Complementary to these are several spectral indices which we provide in Table \ref{tab:specInd} with equivalent spectral types and uncertainties estimated using a Monte-Carlo random sampling approach. The weighted average spectral type from the indices is L1.1$\pm$0.3. The 2MASS $J-Ks=1.33$ colour of \textbeta~Cir~B is very close to the mean value for field gravity L1 dwarfs in figure 1 of \citet{faherty13}. Using the more accurate Hipparcos parallax of the primary, the absolute 2MASS $J$, $H$, and $K_{\text{s}}$ magnitudes for \textbeta~Cir~B are $12.11\pm0.06$, $11.25\pm0.04$, and $10.78\pm0.04$ respectively. These all fall between the mean values for L1 and L1.5 dwarfs in \citet{dupuy12} table 16, which adds weight to the spectral type and rules out an unresolved binary with components of similar spectral types. 
We adopt a spectral type of L1.0$\pm$0.5 based on the combination of the spectral standard comparison and spectral indices.

\begin{table}
\caption{Spectral indices of \textbeta~Cir~B .}
\label{tab:specInd}
\begin{center}
\begin{tabular}{lccc}
\hline
Index & Value & Spectral Type & Reference\\
\hline
H$_2$O J & $0.948\pm0.004$ & L0.4$\pm$0.8 & 1 \\
H$_2$O H & $0.832\pm0.002$ & L1.3$\pm$1.0 & 1 \\
CH$_4$ J & $0.886\pm0.002$ & L1.4$\pm$0.7 & 1 \\
CH$_4$ K & $1.005\pm0.001$ & L4.2$\pm$1.1 & 1 \\
H$_2$O   & $1.204\pm0.004$ & L0.8$\pm$0.7 & 2 \\
H$_2$O D & $0.987\pm0.003$ & M9.8$\pm$0.8 & 3 \\
H$_2$O 1 & $0.684\pm0.004$ & L0.9$\pm$1.6 & 4 \\
H$_2$O 2 & $0.840\pm0.002$ & L1.4$\pm$1.6 & 4 \\
\hline
\end{tabular}
\end{center}
$^1$ \citet{burgasser07}\\
$^2$ \citet{allers07}\\
$^3$ \citet{mclean03}\\
$^4$ \citet{slesnick04}
\end{table}

\subsection{Radial Velocity}\label{RV}
We measured the radial velocity of \textbeta~Cir~B by comparison of the $1.1-1.3$~\textmu{m} and $2.20-2.35$~\textmu{}m regions of the spectrum to model spectra for a range of brown dwarfs with different T$_{\text{eff}}$ and log g. Radial velocities for the $1.1-1.3$~\textmu{m} and $2.20-2.35$~\textmu{}m regions were $9.69\pm1.71$~km~s$^{-1}$ and $9.78\pm0.78$~km~s$^{-1}$ respectively, relative to the Local Standard of Rest. Their close agreement with each other gives us confidence in their reliability and the similarity to the existing radial velocity measurement of \textbeta~Cir~A ($9.6\pm2.0$~km~s$^{-1}$ \citealt{kharchenko07}) lends further weight to the argument of companionship. The inverse variance weighted average radial velocity of \textbeta~Cir~B is $9.76\pm0.71$~km~s$^{-1}$.

\subsection{Surface Gravity}
Brown dwarfs contract as they age, evolving from low to high surface gravity, and it has long been known certain features in their spectra are gravity-sensitive (e.g. \citealt{steele95}, \citealt{lucas01}, \citealt{mcgovern04}, \citealt{faherty12}, \citealt{allers13}, \citealt{canty13}). In the near infrared, the strengths of FeH, VO, Na I, and K I absorption features and collisionally induced H$_2$ absorption in the $H$ and $K$ bands have been shown to be good tracers of surface gravity.
These features can be quantitatively assessed using the indices and pseudo-equivalent widths (EWs) presented by \citet{allers13}, and by visual comparison to objects of a similar spectral type and known surface gravity. The indices and EWs for \textbeta~Cir~B are given in Tables \ref{tab:loggInd} and \ref{tab:EW} respectively.

\begin{figure}
\begin{center}
\begin{tabular}{c}
\epsfig{file=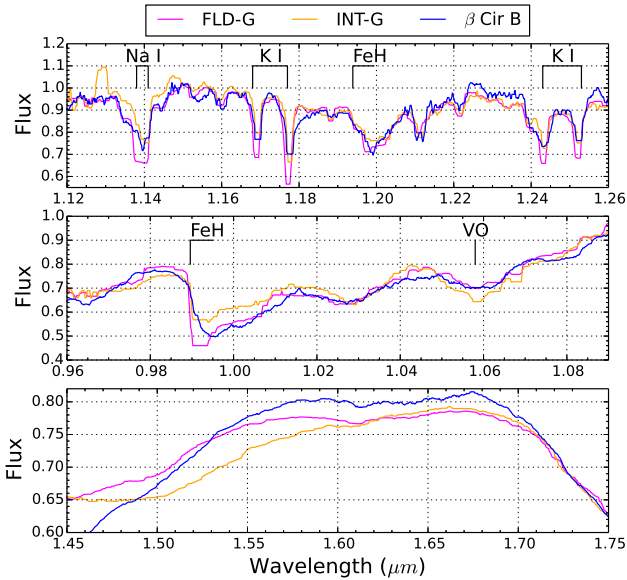,width=1\linewidth,clip=}
\end{tabular}
\caption{The spectrum of \textbeta~Cir~B compared to the field and intermediate gravity L0 dwarfs 2MASS~J17312974+2721233 and 2MASS~J15474719-2423493 respectively. The spectra have been smoothed to approximate the same resolution but to a different degree in each panel. 
\textit{Upper panel}: The $J$ band spectra have been normalised to the integrated flux of \textbeta~Cir~B in the continuum regions of the FeH$_J$ index defined in \citet{allers13}.
\textit{Middle panel}: The $z$ band spectra have been normalised to the integrated flux of \textbeta~Cir~B in the continuum regions of the FeH$_z$ and VO$_z$ indices defined in \citet{allers13}.
\textit{Bottom panel}: The $H$ band spectra have been normalised to the integrated flux of \textbeta~Cir~B in the regions of $H$ band continuum defined in \citet{allers13}.}
\label{fig:gravFeats}
\end{center}
\end{figure}

The strength of FeH absorption is quantified using the FeH$_z$ and FeH$_J$ indices which test the features at $0.99$~\textmu{}m and $1.20$~\textmu{}m respectively. The FeH$_z$ index suggests strong FeH absorption at $0.99$~\textmu{}m and visual inspection confirms this, suggestive of field gravity. The FeH$_J$ index suggests the feature at $1.20$~\textmu{}m is somewhat weaker, suggestive of intermediate gravity, though this is not obvious on visual inspection (see Figure \ref{fig:gravFeats}).

VO absorption is known to be stronger in low surface gravity objects. The VO$_z$ index quantifies the strength of VO absorption at $1.06$~\textmu{}m and has proven to be an excellent tracer of surface gravity in early L dwarfs. The VO$_z$ index of \textbeta~Cir~B is suggestive of field gravity. Visual inspection confirms that there is relatively weak VO absorption at $1.06$~\textmu{}m.

The Na I and K I alkali lines in the $J$ band are weaker in low surface gravity objects. The EWs of these lines can be measured for moderate resolution spectra. Table \ref{tab:EW} gives the EWs for the $J$ band alkali lines of \textbeta~Cir~B. The K I line values are close to the intermediate gravity threshold given by \citet{allers13} but formally are not precise enough to confidently classify the object. The strength of these lines in \textbeta~Cir~B generally appear comparable to the \textsc{int-g} object in Figure \ref{fig:gravFeats}. The Na I line EW is suggestive of intermediate gravity according to \citet{allers13}, though it appears somewhat stronger than the \textsc{int-g} object in Figure \ref{fig:gravFeats} if not quite as strong as the \textsc{fld-g} object.

Finally, the shape of the $H$ band continuum is a well known indicator of low surface gravity. Collisionally induced H$_2$ absorption and FeH absorption reduce flux in the $H$ band as an object evolves to higher surface gravity \citep{borysow97}, this manifests as an apparent flattening of the peak of the $H$ band. A flat $H$ band is clearly visible in the spectrum of \textbeta~Cir~B (see Figures \ref{fig:L1_comparison} and \ref{fig:gravFeats}). The "shoulder", an increase in flux at $\sim1.57$~\textmu{}m seen in the $\sim50-150$~Myr L0 dwarf 1RXS J2351+3127 B and the $\sim200-400$~Myr M9 dwarf LP~944-20 (\citealt{bowler12}, figure 8) is not apparent. The H-cont index measurement echoes this, falling outside the bounds containing the intermediate gravity dwarfs.

Following the \citet{allers13} classification scheme for moderate resolution spectra, the gravity class is taken as the median of: the lowest gravity of the FeH indices (\textsc{int-g}); the VO$_z$ index (\textsc{fld-g}); the mean of the alkali line EWs (inconclusive); and the H-cont index (\textsc{fld-g}). The median in this instance is \textsc{fld-g}.

The \citet{canty13} H$_2$($K$) index for \textbeta~Cir~B is $1.044\pm0.003$, this is typical of field objects for its spectral type (see also \citealt{schneider14}).

\begin{table}
\caption{\citet{allers13} surface gravity indices and equivalent scores of \textbeta~Cir~B. Note that the K I$_J$ index is not used to determine a gravity class using medium resolution spectra.}
\label{tab:loggInd}
\begin{center}
\begin{tabular}{lcc}
\hline
Index & Value & Gravity score\\
\hline
FeH$_z$  & $1.377\pm0.011$ & \textsc{fld-g}\\
FeH$_J$  & $1.210\pm0.007$ & \textsc{int-g}\\
VO$_z$   & $1.094\pm0.004$ & \textsc{fld-g}\\
K I$_J$  & $1.148\pm0.002$ & \textsc{int-g}\\
H-cont   & $0.892\pm0.001$ & \textsc{fld-g}\\
\hline
\end{tabular}
\end{center}
\end{table}

\begin{table}
\caption{\textbeta~Cir~B alkali line pseudo-equivalent widths and gravity scores from \citet{allers13}. Note that the K I line at 12437\AA~is blended with an FeH feature and is not used for the surface gravity analysis. A question mark in the gravity score column indicates that the equivalent width is not precise enough to classify the object.}
\label{tab:EW}
\begin{center}
\begin{tabular}{lccc}
\hline
Feature & Wavelength & Equivalent Width & Gravity \\
        &   {\AA}     &    {\AA}        & score\\
\hline 
Na I& 11396 & $11.0\pm1.7$ & \textsc{int-g} \\
K I & 11692 & $5.9\pm1.5$  & ? \\
K I & 11778 & $9.3\pm0.8$  & ? \\
K I & 12437 & $6.5\pm0.9$  & ... \\
K I & 12529 & $6.0\pm1.5$  & ? \\
\hline
\end{tabular}
\end{center}
\end{table}

\subsection{Mass estimate}
Interpolation over the $200-600$~Myr \citet{baraffe15} isochrones and using the values for age and T$_{\text{eff}}$ from Table \ref{tab:betCirProperties} gives a mass and $\log{g}$ for \textbeta~Cir~B of $0.056\pm0.007$~M$_{\odot}$ 
and $5.15\pm0.04$ respectively. Note that these uncertainties do not incorporate the uncertainties inherent to the model.

\subsection{Variability}
To check for variability of \textbeta~Cir~B in the $K_{\text{s}}$ band we used standard catalogue \textit{aperMag2} photometry of the 58 observations between 5 March 2010 and 3 July 2013 that met our astrometric criteria.
These are data from both pawprints of VVV tile d018 that cover \textbeta~Cir~B. We corrected the magnitudes using 15 background reference sources from within $1$\arcmin~with the same $K_{\text{s}}$ magnitude to within $\pm0.5$~mags and no obvious source within $2$\arcsec. The uncertainties on the corrected magnitudes we took as the formal magnitude error from the catalogue added in quadrature to the RMS scatter of the offsets of the 15 reference sources.
Magnitudes from pairs of observations taken within 30 minutes were averaged using inverse variance weighting.
There is no sign of significant variability of \textbeta~Cir~B above an RMS level of $0.013$~mag ($\approx1.2\%$). A non-detection of variability above this level is typical for early L type dwarfs (see e.g. \citealt{radigan14}).

\section{Discussion and Summary}\label{discussion}

We have discovered a new benchmark L dwarf companion to a nearby A3V star with an age of $370-500$~Myr. A range of properties of the \textbeta~Cir~system are given in Table \ref{tab:betCirProperties}. The projected physical separation of $\sim$6700~au is amongst the widest 10\% known for brown dwarf companions, see \citealt{deacon14} and the mass ratio of the pair ($q \approx 0.028$) is also unusually small (see figure 11 of \citealt{derosa14b}). 
Measurements of individual chemical abundances exist for \textbeta~Cir~A \citep{erspamer03}, but as noted in Section \ref{age} these cannot be assumed to reflect the true abundances of either component of the system. The scatter in values for individual abundances, and the apparently normal field brown dwarf spectrum of \textbeta~Cir~B are evidence that the measured highly supersolar abundances of V and Na ([V/H]~$=0.52$, [Na/H]~$=0.74$) do not in fact reflect the composition of the brown dwarf. Instead the normal L dwarf spectrum is consistent with the approximately solar metallicity that we derived in Section \ref{age} for the primary by fitting its location on the Hertzsprung-Russell diagram. (If valid, the measured abundances would have been expected to push the gravity sensitive VO and NaI features towards lower gravity and higher gravity respectively).

\textbeta~Cir~B is located in a useful but sparsely occupied part of the age and temperature grid for brown dwarfs where surface gravities are predicted to approach those of mature field objects. Near infrared signatures of low gravity are visible in the Pleiades at $125$~Myr \citep{bihain10} and in other L dwarf benchmarks with ages in the range $50-200$~Myr, e.g. G~196-3B (\citealt{rebolo98}; \citealt{allers07}), CD-35 2722B \citep{wahhaj11} and 1RXS J2351+3127 B \citep{bowler12}. By contrast, the somewhat older L4+L4 pair HD~130948~BC, aged $790\substack{+220 \\ -150}$~Myr, was described by \citet{allers10} as the youngest L dwarf benchmark showing no signs of low surface gravity. Similarly, the recently discovered L5 companion \textzeta~Del~B, at $525\pm125$~Myr \citep{derosa14b}, showed no signs of low gravity. The L4.5+L6 binary Gl~417~BC is also interesting in this context. The discovery paper \citep{kirkpatrick01} assigned a most likely age of $80-300$~Myr and found only marginal evidence for low gravity in the optical spectrum (see also \citealt{kirkpatrick08}). \citet{allers10} found that the near infrared spectrum of the pair also indicates normal field gravity but they find a gyrochronal age of $750\substack{+140 \\ -120}$~Myr for the primary. Moreover, the preferred age given by \citet{kirkpatrick01} was based on three quantitative estimates for the primary: $80-250$~Myr from X-ray activity, $150-400$~Myr from gyrochronal calculations and $\sim600$~Myr from Li abundance. Given the diversity of these ages it is difficult to use the Gl~417 system to put a figure on the age at which low gravity signatures disappear.  
At the age of the Pleiades ($\sim120$~Myr) the surface gravity of a $2100$~K L-type dwarf is predicted to be $0.3$~dex lower than at $400$~Myr \citep{baraffe15}. However, the predicted gravity is only $0.2$~dex higher for a (more massive and smaller) $5$~Gyr object than a $400$~Myr object with the same temperature.

While we cannot place too much emphasis on one object, the absence of any compelling indication of low gravity in our high quality intermediate resolution spectrum suggests that low gravity signatures disappear from near infrared spectra of early L dwarfs by the age of $\sim370-500$~Myr.

\begin{table}
\caption{Properties of the \textbf{\textbeta}~Cir~system.}
\label{tab:betCirProperties}
\begin{center}
\begin{tabular}{lccl}
\hline
   				& \textbeta~Cir~A 	& \textbeta~Cir~B		&		\\
\hline
  Right Ascension$^a$		& 15h17m30.85s		& 15h17m21.60s		&		\\
  Declination$^a$		& -58$^{\circ}$48\arcmin04.34\arcsec	
  							& -58$^{\circ}$51\arcmin30.0\arcsec	
  										&		\\
  Parallax 			& $32.73\pm0.19$$^b$	& $33.6\pm1.6$		& mas		\\
  $\mu_{\alpha}\cos\delta$ 	& $-97.4\pm0.28$$^b$ 	& $-96.7\pm2.5$$^c$		& mas~yr$^{-1}$	\\
  $\mu_{\delta}$		& $-134.15\pm0.22$$^b$	& $-133.1\pm2.6$$^c$	& mas~yr$^{-1}$	\\
  Radial Velocity		& $9.6\pm2.0$$^d$	& $9.76\pm0.71$	& km~s$^{-1}$	\\
  Spectral Type 		& A3V$^e$		& L1.0$\pm$0.5		&		\\
  2MASS J			& $3.93\pm0.25$ 	& $14.54\pm0.06$	& mag		\\
  2MASS H			& $3.81\pm0.24$		& $13.68\pm0.04$	& mag		\\
  2MASS Ks			& $3.88\pm0.18$		& $13.21\pm0.04$	& mag		\\
  VVV Z\footnotemark[1] &                   & $16.7\pm0.1$  & mag     \\
  VVV Y\footnotemark[1] &                   & $15.6\pm0.1$  & mag     \\
  VVV J             &                   & $14.41\pm0.02$  & mag     \\
  VVV H             &                   & $13.70\pm0.02$  & mag     \\
  VVV Ks            &                   & $13.16\pm0.02$  & mag     \\
  T$_{\text{eff}}$	& $8676\pm33$		& $2084\pm150$$^f$	& K		\\
  log~g			& 	$4.21$$^g$		& 	$5.15\pm0.04$$^h$	& dex		\\
  Age				&\multicolumn{2}{c}{$370$ to $500$}			& Myr		\\
  Mass				& $1.96\substack{+0.03 \\ -0.01}$		& 		$0.056\pm0.007$$^h$		& M$_{\odot}$\\
  Angular Separation$^a$	& \multicolumn{2}{c}{$217.8$} 			& \arcsec	\\
  Projected Separation$^a$	& 	\multicolumn{2}{c}{$6656$}		& au		\\
\hline
\end{tabular}
\end{center}
$a$ - Epoch J2000.0; $b$ - \citet{vanleeuwen07}; $c$ - inclusive of the relative to absolute correction derived from PPMXL; $d$ - \citet{kharchenko07}; $e$ - \citet{gray06}; $f$ - using \citet{marocco13} equation 6; $g$ - \citet{erspamer03}; $h$ - based on interpolation of the \citet{baraffe15} models, the stated errors do not include uncertainties inherent to the models.
\end{table}

\footnotetext[1]{The uncertainties given for the VVV photometry are the calibration uncertainties, which are the main source of error for this relatively bright source. These are worst for the $Z$ and $Y$ band data, owing to interstellar extinction in the Galactic plane towards 2MASS calibrator stars. Improvement is expected from a recently begun VVV $ZY$ calibration campaign.} 

\section*{Acknowledgments}
We are grateful to K. Allers for supplying the low gravity BD spectra. LS is grateful to N. Deacon and M. Irwin for helpful viva comments, and R. Smart for various useful discussions.
We acknowledge use of data from the ESO Public Survey programme ID 179.B-2002 taken with the VISTA telescope, data products from CASU, and funding from the FONDAP Center for Astrophysics 15010003, the BASAL CATA Center for Astrophysics and Associated Technologies PFB-06, the FONDECYT from CONICYT.
This research uses data gathered with the 6.5 meter Magellan Telescopes located at Las Campanas Observatory, Chile.
LS acknowledges a studentship funded by the Science \& Technology Facilities Research Council (STFC) of the UK; PWL acknowledge the support of a consolidated grant (ST/J001333/1) also funded by STFC.
Support for RK, DM, MG, VV, CCP, JB, and JCB is provided by the Ministry of Economy, Development, and Tourism’s Millennium Science Initiative through grant IC120009, awarded to The Millennium Institute of Astrophysics (MAS), they also acknowledge CONICYT REDES No. 140042 project. RK and DM are supported by Fondecyt Reg. No. 1130140 and No. 1130196, respectively.
MG acknowledges support from Joined Committee ESO and Government of Chile 2014 and  Fondecyt  Regular No. 1120601.
This research has made use of: The SIMBAD database and VizieR catalogue access tool, operated at CDS, Strasbourg, France; NASA's Astrophysics Data System Bibliographic Services; and the SpeX Prism Spectral Libraries, maintained by Adam Burgasser at \href{http://pono.ucsd.edu/~adam/browndwarfs/spexprism}{pono.ucsd.edu/$\sim$adam/browndwarfs/spexprism}.

\bibliographystyle{mn2e}
\bibliography{main}

\end{document}